\documentclass[journal]{vgtc}                     

\onlineid{3766}

\vgtccategory{Research}

\vgtcpapertype{please specify}

\title{\emph{InsightLens}: Augmenting LLM-Powered Data Analysis with Interactive Insight Management and Navigation}

\author{%
  Luoxuan Weng,
  Xingbo Wang,
  Junyu Lu,
  Yingchaojie Feng,
  Yihan Liu,
  Haozhe Feng,
  Danqing Huang, and 
  Wei Chen
}

\authorfooter{
  \item
  	Luoxuan Weng, Junyu Lu, Yingchaojie Feng, Yihan Liu, and Wei Chen are with the State Key Lab of CAD\&CG, Zhejiang University. E-mail: \{lukeweng, junyulu, fycj, liuyihan1024, chenvis\}@zju.edu.cn.
  \item
  	Xingbo Wang is with Weill Cornell Medical College, Cornell University. E-mail: xingbo.wang@med.cornell.edu.
   \item 
        Haozhe Feng and Danqing Huang are with Tencent Inc. E-mail: \{aidenhzfeng, daisyqhuang\}@tencent.com.
}

\abstract{
The proliferation of large language models (LLMs) has revolutionized the capabilities of natural language interfaces (NLIs) for data analysis. LLMs can perform multi-step and complex reasoning to generate data insights based on users' analytic intents. However, these insights often entangle with an abundance of contexts in analytic conversations such as code, visualizations, and natural language explanations. 
This hinders efficient recording, organization, and navigation of insights within the current chat-based LLM interfaces. 
In this paper, we first conduct a formative study with eight data analysts to understand their general workflow and pain points of insight management during LLM-powered data analysis.
Accordingly, we introduce \name, an interactive system to overcome such challenges. Built upon an LLM-agent-based framework that automates insight recording and organization along with the analysis process, \name \space visualizes the complex conversational contexts from multiple aspects to facilitate insight navigation.
A user study with twelve data analysts demonstrates the effectiveness of \name, showing that it significantly reduces users' manual and cognitive effort without disrupting their conversational data analysis workflow, leading to a more efficient analysis experience.
}


\teaser{
  \centering
  \includegraphics[width=0.98\linewidth, alt={The user interface of \textit{InsightLens}.}]{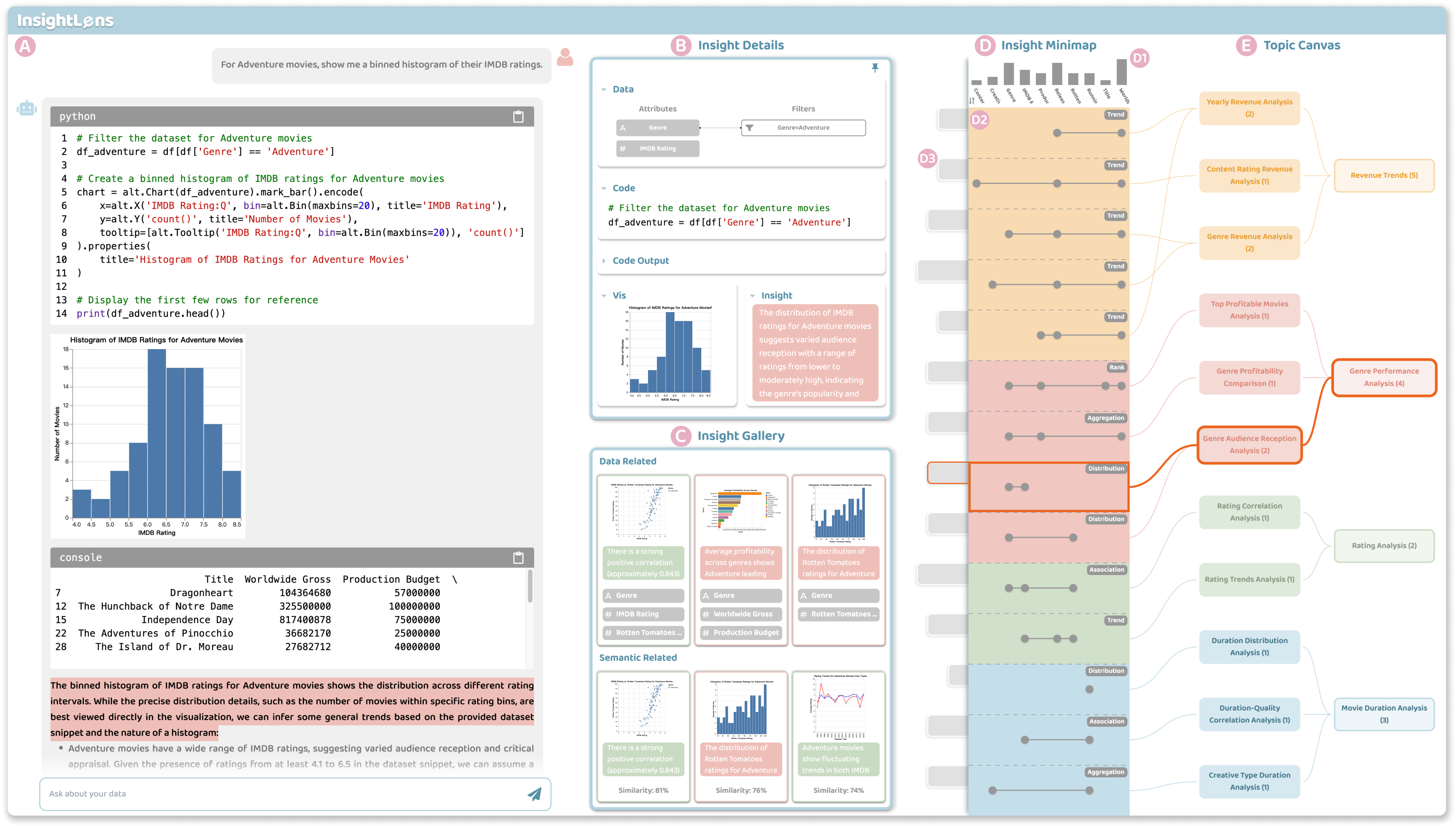}
  \caption{%
  	The user interface of \name. The \textit{Chat Window} (A) enables conversational interactions between users and LLMs. The \textit{Insight Details} (B) displays the currently focused insight's summary with its relevant data context and supporting evidence. The \textit{Insight Gallery} (C) presents the corresponding related insights in terms of data and semantics. The \textit{Insight Minimap} (D) visualizes the analysis process chronologically based on each insight. The \textit{Topic Canvas} (E) provides the hierarchical topic structure of all insights.
  }
  \label{fig:system}
}

\graphicspath{{figs/}{figures/}{pictures/}{images/}{./}} 

\usepackage{tabu}                      
\usepackage{booktabs}                  
\usepackage{lipsum}                    
\usepackage{mwe}                       

\usepackage{mathptmx}                  

\usepackage{url}

\usepackage{multicol}
\usepackage{rotating}
\usepackage{multirow}

\newcommand{\name}{\textit{InsightLens}}

\usepackage{xspace,xpunctuate}
\usepackage{tabularx}

\newcommand{\ie}{\textit{i.e.},\xspace}
\newcommand{\etal}{\xspace\textit{et al.}\xspace}
\newcommand{\eg}{\textit{e.g.},\xspace}
\newcommand*{\icon}[1]{\scalerel*{\includegraphics{#1}}{\strut}}

\usepackage{enumitem}

\begin{document}

\maketitle

\section{Introduction}

Natural language interfaces (NLIs) for data analysis~\cite{9222342, cox2001multi} have received much attention in recent years. Users express their analytic intents and data-related questions in natural language (NL), prompting NLIs to generate corresponding results or visualizations for further analysis. Recently, large language models (LLMs), such as GPT-4~\cite{achiam2023gpt} and LLaMA~\cite{touvron2023llama}, have achieved unprecedented performance in NL understanding, reasoning, and generation.
They have become the backbones for NLIs (\eg ChatGPT's Advanced Data Analysis~\cite{openai}) to enhance conversational data analysis~\cite{xie2023openagents, gu2023data}, hereafter referred to as \textit{LLM-powered data analysis}.

During LLM-powered data analysis, LLMs can perform multi-step and complex reasoning to derive data insights based on users' queries about the dataset and the previous conversational contexts~\cite{shen2024datastoryautomaticanimated}.
This process also generates various intermediate outputs, such as code, visualizations, and NL explanations~\cite{chopra2023conversational}.
In a typical round of question and answer (Q\&A) within analytic conversations, users must carefully examine and understand the insights generated by LLMs, which are usually entangled with an abundance of intermediate outputs.
Furthermore, data analysis is an exploratory and iterative procedure that commonly involves multiple rounds of Q\&A. As such, maintaining awareness and keeping track of the entire analyses is essential for making informed decisions and determining future exploration directions~\cite{wang2022interactive, 10.1145/3472749.3474792}. 
This emphasizes the need to \textit{record}, \textit{organize}, and \textit{navigate} the insights generated throughout the analysis process.

However, recording, organizing, and navigating insights within the current chat-based LLM interfaces is tedious and inefficient, especially given the intertwined data and semantic context involved.
During data analysis, insights need to be recorded with their supporting evidence (\ie intermediate outputs like visualizations) for sharing and reporting purposes~\cite{10.1145/3491102.3517485}. This requires users to navigate back and forth in the conversation to locate the needed information. As analytic conversations are usually lengthy and overwhelmed with various contexts, this process often causes significant manual effort.
Existing tools primarily focus on tracking the provenance of a single form of context
(\eg data~\cite{10301695}, code~\cite{10.1145/3290605.3300322}, or visualization~\cite{8788592}) and are not tailored for conversational interfaces. 
This limitation impedes efficient insight understanding and recording that involves multiple forms of context in the conversation.
The situation is exacerbated for insight organization and navigation.
Given the increasing volume of LLM-generated insights and the quickly expanding conversation length,
users face a substantial cognitive load. They struggle to manage and organize these insights efficiently in a structured and readable manner, while also maintaining convenient navigation.
Although numerous systems have emerged to help users organize and explore LLMs' responses in various scenarios~\cite{suh2023structured, 10.1145/3586183.3606737, 10.1145/3586183.3606756}, they often fall short in addressing the challenges in data analysis conversations. 
Some studies focus only on the semantic context (\eg topic changes~\cite{liang2023c5}) and ignore the data context~\cite{10.1145/3472749.3474792, 8019833}. Others focus on monitoring and verifying single rounds of Q\&A~\cite{waitgpt, 10.1145/3654777.3676345}, which is insufficient to comprehensively explore the entire analysis process containing multi-round Q\&A.

Therefore, our goal is to make conversational data analysis more trackable and navigable for users, and to support on-the-fly recording and organization of insights through a new interaction paradigm. Informed by a formative interview study with eight experts of LLM-powered data analysis, we summarize the challenges of existing chat-based LLM interfaces for data analysis. Accordingly, we present \name \space, an interactive system to facilitate insight recording, organization, and navigation. Rather than burdening users with manually managing insights from the complex conversational contexts, \name \space adopts an LLM-agent-based framework for automatic recording and organization of insights during conversational data analysis.
Moreover, \name \space augments traditional chat-based interfaces with multi-level and multi-faceted visualizations to aid in monitoring and navigating the entire conversation. Specifically, it features an \textit{Insight Minimap} and a \textit{Topic Canvas} that progressively evolve along with the analysis process to reveal the temporal shifts of data and semantic context.
They provide on-the-fly feedback to guide data exploration without disrupting the conversational workflow.
To evaluate the effectiveness of \name, we conducted a technical evaluation and a user study. The technical evaluation demonstrated a satisfactory performance of the agent-based framework in accurately recording and organizing insights.
The user study revealed that the system can significantly reduce users' manual and cognitive effort for insight management and navigation in LLM-powered data analysis, leading to an improved analysis experience.

In summary, the major contributions of our work are:

\begin{itemize}[leftmargin=*]
\vspace{-1mm}
\item
A formative study that identifies critical challenges and summarizes design requirements for insight management and navigation during LLM-powered data analysis.

\vspace{-1mm}
\item
\name, a system that facilitates insight recording, organization, and navigation through a novel LLM-agent-based framework and interactive visualizations.

\vspace{-1mm}
\item
A technical evaluation and a user study that demonstrate the effectiveness of \name.

\end{itemize}
\section{Related Work}

\subsection{NLIs for Data Analysis}

Natural language is an intuitive modality for data interaction, significantly lowering the barriers of data analysis~\cite{8019833}. Therefore, NLIs for data analysis have been extensively studied in multiple fields including databases~\cite{affolter2019comparative}, NLP~\cite{liu-etal-2023-jarvix}, and visualization~\cite{10.1145/2807442.2807478}. Chen\etal~\cite{10.1145/3491102.3517485} divided these systems into two types: NLIs for data queries and for visualizations. Following this categorization, we review previous works and discuss recent advancements in LLM-powered data analysis.

\textbf{NLIs for data queries} convert NL utterances into machine-readable formats like SQL and Python to execute on knowledge bases~\cite{Binder}. Early systems relied on pattern-matching~\cite{10.1145/3132847.3132977}, parsing strategies~\cite{10.14778/2994509.2994536}, or rule-based methods~\cite{10.1007/978-3-319-34129-3_19} to understand the semantic structures of queries~\cite{affolter2019comparative}. Later, neural approaches~\cite{wang-etal-2020-rat, guo-etal-2019-towards} trained end-to-end networks to directly generate executable SQL queries from NL inputs, addressing issues like ambiguities or fuzzy linguistic coverage. Recently, training-free strategies using LLMs have emerged, achieving state-of-the-art performance~\cite{zhou-etal-2022-tacube} by leveraging LLMs' reasoning abilities with minimal in-context examples, as demonstrated by systems like Binder~\cite{Binder}.

\textbf{NLIs for visualizations (V-NLIs)}~\cite{8019860, 8440860} take a step further by generating visualizations based on query results. Introduced by Cox\etal~\cite{cox2001multi}, these systems enable users to focus more on their data rather than manipulating complex visual interfaces. Many efforts aim to resolve ambiguities or underspecifications in input queries~\cite{10.1145/3301275.3302270, 10.1145/2807442.2807478}. For example, NL4DV~\cite{9222342} explicitly highlighted ambiguities in its generated visualization specifications. Other research explores analytic context to maintain a conversational flow~\cite{8933766, 8986918}. Evizeon~\cite{8019833} applied pragmatics principles and defined context transition types (\ie \textit{continue}, \textit{retain}, and \textit{shift}). Based on this, Snowy~\cite{10.1145/3472749.3474792} recommended context-aware utterances for conversational visual analysis. Similarly, our work also highlights data context transitions during analysis.

Recently, \textbf{analytical assistants powered by LLMs} have become a prevalent paradigm~\cite{10.1145/3544548.3580940, waitgpt}.
Many commercial business intelligence (BI) platforms like Power BI~\cite{powerbi} and Tableau~\cite{tableau} have integrated LLM support for chat-based insight discovery and dashboard generation.
Empirical studies have explored conversational challenges~\cite{chopra2023conversational} and user behaviors~\cite{gu2023data} during LLM-powered data analysis.
Automated LLM-based tools have also been developed, such as InsightPilot~\cite{ma-etal-2023-insightpilot} for simplifying data exploration by generating insights and AI Threads~\cite{hong2023conversational} for creating and refining charts through a multi-threaded chatbot.

Overall, the extensive studies on NLIs for data analysis provides a solid foundation for our work.
We focus on LLM-powered data analysis for its recent prevalence and rather immature interaction schemes~\cite{gu2023analysts}.
While conversations are natural and intuitive, this new paradigm brings unique challenges that increase manual and cognitive load on users~\cite{10.1145/3654777.3676345}. For example, recent studies have explored cognitive issues like tedious code verification~\cite{waitgpt} and overwhelming response comprehension~\cite{chopra2023conversational}. In this work, we aim to identify the pain points in conversational data analysis and help users better manage insights along the way.

\subsection{Analytic Provenance in Data Analysis}

Analytic provenance tracks the history and evolution of various analytic context, such as data~\cite{10.1145/3411764.3445063} and visualizations~\cite{8788592}, which helps users better understand the analysis process. Ragan\etal~\cite{7192714} introduced an organizational framework to characterize different types and purposes of provenance, which Madanagopal\etal~\cite{8788592} further expanded by mapping tasks to provenance types. Researchers have also proposed various techniques for effective provenance management~\cite{7194834} and presentation~\cite{9768153}.
For example, Berant\etal~\cite{8731342} used cell-based provenance with NL utterances to explain queries over data tables, while DIY~\cite{10.1145/3397481.3450667} enabled users to evaluate NLIs' correctness on databases by visualizing data subset transformations. XNLI~\cite{10026499} provided interactive widgets to depict visualization provenance in V-NLIs for explanation and diagnosis. More recently, WaitGPT~\cite{waitgpt} visualized the step-by-step generation of data code to help users monitor and verify LLM-powered data analysis. Our work extends these efforts by extracting and tracking insights along with other analytic context, binding these insights with relevant evidence (\eg visualizations) to enhance user comprehension.

\subsection{Exploration of LLM Responses}

Limitations of the linear conversational structures pose challenges in supporting complex information tasks with LLMs~\cite{liu2023sprout}. Therefore, numerous visual interfaces have been introduced to facilitate LLM response exploration~\cite{hoque2024hallmark, lu2024agentlens}. For example, Sensecape~\cite{10.1145/3586183.3606756} supported multi-level exploration and sensemaking, while Graphologue~\cite{10.1145/3586183.3606737} created interactive diagrams based on named entity recognition, both enhancing users' understanding of individual responses. Luminate~\cite{suh2023structured} further supported structured examination of multiple responses by generating a multi-dimensional design space for human-AI co-creation. Additionally, C5~\cite{liang2023c5} and Memory Sandbox~\cite{10.1145/3586182.3615796} addressed conversational context management issues by visualizing topic transitions and enabling transparent memory management, respectively. However, these interfaces are not tailored for data analysis, limiting their effectiveness in managing data insights. Our work extends this research by offering multi-level and multi-faceted visualizations to facilitate insight management and navigation in conversational data analysis.
\section{Formative Study}

The target users of our system are data analysts who utilize LLMs for analytical tasks.
To understand the pain points and challenges of existing chat-based interfaces, we conducted a formative interview study.
This study specifically examined how participants record, organize, and navigate insights generated by LLMs during conversational data analysis within a ChatGPT-like interface.
Based on our findings, we derived four design requirements to facilitate insight management.

\subsection{Participants and Procedure}

\textbf{Participants.}
Eight data analysts from various domains, including finance and e-commerce were interviewed (E1-8, 3 females and 5 males, age from 25 to 32). Four participants were senior data analysts, while the remaining four were juniors or intermediates. All of them had recently used LLMs for generating data visualizations or insights.

\textbf{Settings.} We created an analytical chatbot based on Open Interpreter~\cite{openinterpreter} with GPT-4, akin to ChatGPT's Advanced Data Analysis. It could be prompted with queries to generate code for data processing and visualization, and then interpret execution results to derive insights.

\textbf{Procedure.} Participants were asked to perform open-ended data analysis~\cite{10026499} with the system to explore the movies dataset from Vega,
which contains 709 rows and 10 columns. Similar to their daily work, the task was to derive and record data insights based on the dataset and produce a clear, structured report.
We collected their feedback on the analysis experience, focusing on how they acquired information from the conversation and organized and navigated the insights for summarization or further exploration. We then identified challenges and obstacles they encountered. The interviews were conducted online and lasted about 60 to 80 minutes.

\subsection{Findings}

We observed how participants managed the generated insights throughout the analysis process. For each single round of Q\&A, they first reviewed the textual response and visualizations (if any) to grasp the main idea of the message. Some of them then scrolled back to examine the code and its execution results, which were noted as being `\textit{helpful for understanding and reproducibility}' (E5). Subsequently, participants recorded and organized insights through copy-and-paste or screenshots with documentation tools like Google Docs. After collecting enough insights or finishing a specific analytic topic, they navigated previous notes or screenshots to recap findings and plan next steps. However, during the entire process, participants faced several common challenges that decreased analysis efficiency, which are summarized below.

For clarity, we define the terminologies used in the paper.

\begin{itemize}[leftmargin=*]

    \vspace{-1mm}
    \item
    \textbf{Analytic Context:} Properties of the dataset (focused attributes and values), user interactions (analytic intents and data-related questions), intermediate outputs for analytic purposes (code, code outputs, visualizations, and NL explanations), and insights derived by LLMs.

    \vspace{-1mm}
    \item
    \textbf{Insight Evidence:} \textit{Parts} of the intermediate outputs generated by LLMs that \textit{directly} support each insight, including the \textit{specific piece} of code, code outputs, visualizations, and NL explanations.
\end{itemize}

\textbf{C1: Laborious insight recording from overwhelming conversational contexts.}
Recording an insight requires both tediously \textit{`summarizing the key idea of the lengthy response'} (E1) and \textit{`locating relevant information like visualizations as supporting evidence'} (E3). For example, E5 spent much time in scrolling back to copy code snippets and their outputs \textit{`in case of reproducing the results in the future'}. The situation was exacerbated when participants had to iteratively modify their utterances to steer LLMs' behavior, in which case the insight and its evidence would span across multiple responses, causing extra effort for excessive scrolling.
Although LLMs could be explicitly prompted to generate less verbose responses, balancing between comprehensiveness and succinctness was hard to achieve, especially during data analysis. As stated by E4, \textit{`I prefer comprehensive analyses for high-level questions, but only need a quick answer for simple data queries.'}

\textbf{C2: Significant overhead for insight organization.}
Most participants (7/8) organized recorded insights into meaningful subgroups based on data attributes or analytic topics with external documentation tools. This process was described as \textit{`troublesome and painstaking'} (E4), due to the necessity of manually annotating each insight with its characteristics before synthesizing them collectively.
Notably, some participants (3/8) explicitly asked LLMs to help organize insights. However, obtaining satisfactory results required iterative and nuanced prompt engineering, which could disrupt the analysis flow. As stated by E8, \textit{`I had to start another conversation specially used for organization, otherwise the original analysis conversation would become too messy.'}
The frequent switching between different conversation threads and documentation tools was \textit{`frustrating and time-consuming'} (E3). Meanwhile, as the analysis progressed, the document itself became overwhelmed with \textit{`many unordered texts and images'} (E5), which made it even harder for structured organization.

\textbf{C3: Inflexible and inefficient insight browsing and revisiting.}
Participants constantly revisited and navigated previous findings throughout the analysis process. They reported that the lack of \textit{`a high-level insight overview'} (E7) hindered quick navigation and contextual understanding, especially when the conversation became lengthy. The extra cognitive load for insight navigation mainly reflected in two aspects.
First, browsing insights was inconvenient. For example, E3 maintained an outline of her discoveries in Word, but the document soon became lengthy, forcing her to \textit{`repeatedly scroll up and down to browse each section'}, which \textit{`somewhat outweighed the advantages of organizing insights'} (E3). Moreover, participants desired to prioritize significant insights during navigation instead of \textit{`random meandering'} (E4), which was not supported. Second, revisiting previous related insights and their supporting evidence was cumbersome, which is a frequent need during analysis for \textit{`comparison or reference'} (E6) and \textit{`inspiring new discoveries'} (E8), as stated by many participants (5/8). Besides, many participants (5/8) mentioned that they sometimes unknowingly stuck in certain subsets of data attributes (E2, E5) or analytic topics (E1), leading to potential biases. Such issues could have been mitigated if users were \textit{`more aware of the data or semantic changes'} (E1).


\subsection{Design Requirements}

The findings indicate that data analysts struggle with current LLM interfaces for insight management and navigation. To this end, we aim to design a novel interactive system for better recording, organization, and navigation of insights to facilitate a more efficient data analysis experience. The design requirements can be summarized as follows.

\textbf{R1: Support automatic insight recording from LLMs' responses.}
Manual recording of insights from the overwhelming conversation requires users' tedious examination and excessive scrolling (\textbf{C1}). Therefore, the system should constantly monitor the conversation to automatically summarize and record the generated insights and bind relevant insight evidence (\eg code outputs, visualizations) with them, regardless of whether LLMs' responses are verbose or not.

\textbf{R2: Facilitate effective and on-the-fly insight organization.}
Manual organization of insights based on data attributes or analytic topics is inefficient and troublesome (\textbf{C2}), especially when numerous insights and messy analytic context are involved.
Meanwhile, the context switching between different applications or conversation threads incurs extra cognitive load. Hence, the system should organize insights in a non-intrusive manner along with the analysis process.

\textbf{R3: Provide multi-level and multi-faceted insight navigation.}
Browsing and revisiting previous insights from multiple aspects or levels of detail are burdensome (\textbf{C3}). Therefore, the system should support multi-faceted insight navigation (\eg temporal, data attributes, analytic topics). Additionally, insight interestingness~\cite{10.1145/3299869.3314037} and context transitions~\cite{10.1145/3472749.3474792} should be highlighted to help users quickly identify significant insights and enhance analytic comprehensiveness.
To facilitate easier navigation of the entire conversation, an insight-level overview should be provided, with details on demand to inspect each insight with its supporting evidence and other related insights.



\textbf{R4: Adopt familiar and unobtrusive interactions and visual designs for seamless data analysis.} Users generally appreciate the conversational manner for its intuitiveness and user-friendliness.
Therefore, augmenting existing conversational interfaces with seamlessly integrated visualizations is more favorable than creating complex new tools.
To avoid steep learning curves and high switching costs, the system should adopt familiar visual designs and non-intrusive interactions without disrupting the original chat-based workflow.

\begin{figure*}[ht]
  \centering
  \includegraphics[width=\textwidth]{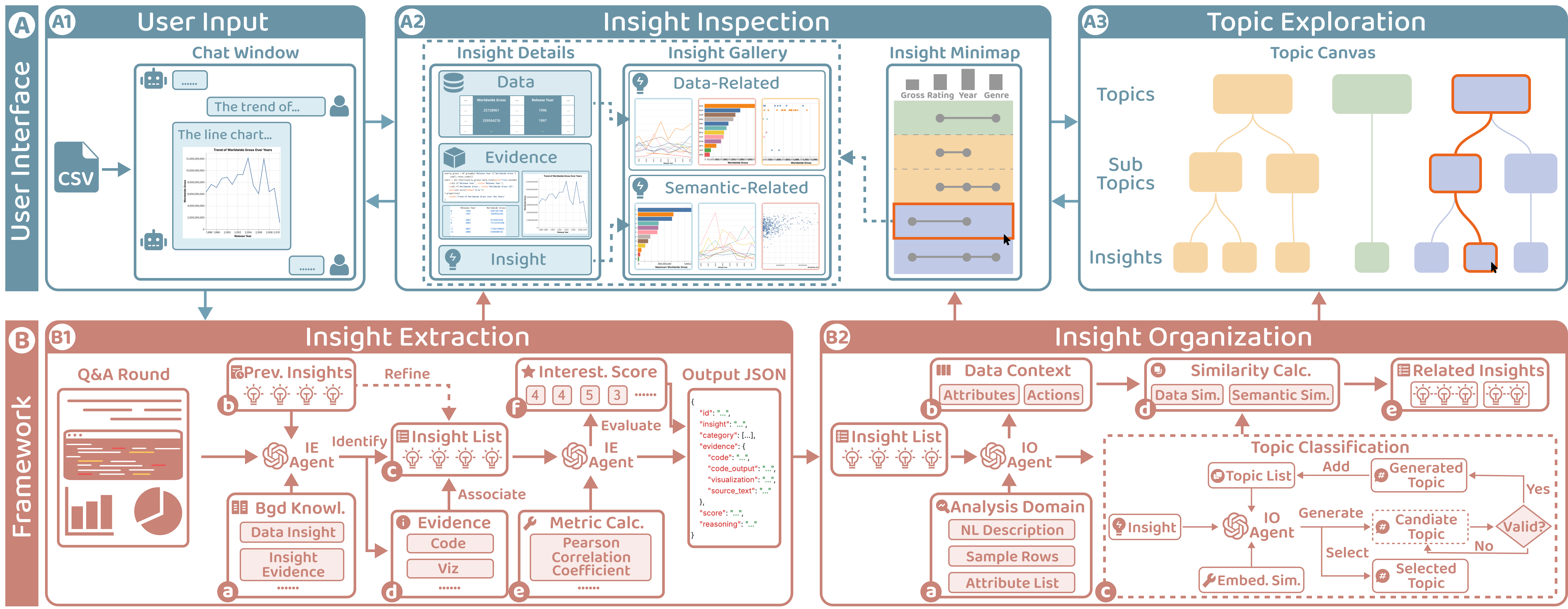}
  \caption{\name \space consists of (A) a user interface and (B) an LLM-agent-based framework.
    While users are (A1) interacting with the analytical chatbot, the \textit{Insight Extraction (IE) Agent} (B1) takes each round of Q\&A for insight extraction and evidence association, as well as interestingness evaluation. Following this, the \textit{Insight Organization (IO) Agent} (B2) organizes the insights by identifying their data context, analytic topics, and related insights. Users can then (A2) inspect the extracted insights and (A3) explore the structured topics with progressively-evolving visualizations.
  }
  \label{fig:workflow}
  \vspace{-5mm}
\end{figure*}

\section{InsightLens: Framework}


Informed by the summarized challenges and design requirements, we propose automating the recording and organization of insights during analysis, and displaying these insights with on-the-fly visualizations to facilitate user navigation. To achieve this, we develop an LLM-agent-based framework (Figure~\ref{fig:workflow}B) that comprises two components: \textit{Insight Extraction (IE)} and \textit{Insight Organization (IO)}, each powered by an LLM-based agent~\cite{xie2023openagents}. The \textit{IE Agent} takes each round of Q\&A within the conversation as input, extracts insights from LLMs' raw responses, and associates them with relevant evidence (\textbf{R1}). It then evaluates the extracted insights' interestingness based on their semantic and statistical significance (\textbf{R3}). These insights are subsequently passed to the \textit{IO Agent}, which examines their data and semantic characteristics and dynamically organizes them along with all previous insights (\textbf{R2, R3}). The framework automatically runs in the background throughout the analysis process without disrupting the conversational workflow (\textbf{R4}).
In this section, we describe the prompt engineering techniques of our framework. Following best practices of designing LLM-based agents, we adopt the ReAct~\cite{yao2022react} paradigm for prompting and equip the agents with specialized tools and in-context memory, allowing them to plan and execute actionable steps to perform various tasks. We use OpenAI's \texttt{gpt-4-0125-preview} model for implementation.



\subsection{Insight Extraction}
\label{sec:ie_agent}

To support automatic insight extraction (\textbf{R1}), the \textit{IE Agent} keeps monitoring the conversation as the analysis progresses (Figure~\ref{fig:workflow}B1).
Upon the user completing one round of Q\&A with the analytical chatbot, the \textit{IE Agent} is responsible for examining the messages and outputting a JSON-formatted insight list. Therefore, the core of the agent design lies in its step-by-step prompt engineering, which is detailed below.

\textbf{Providing background knowledge.} Prior to task delineation, we introduce the definitions of some key terminologies in data analysis such as \textit{insight}, \textit{insight evidence}, and \textit{insight interestingness} (Figure~\ref{fig:workflow}B1(a)), drawing from previous literature~\cite{8805442, 10.1145/3299869.3314037} and our formative study. This allows the agent to be familiar with the essential domain knowledge, facilitating task performance and output quality. Subsequently, we provide a brief description of the dataset currently in play, including its title and attributes. This ensures the agent's focus of the conversation is confined to the content relevant to the data and analytic context, instead of extracting unrelated insights. Finally, we underscore the task and the required output format with a few demonstration examples to better leverage LLMs' in-context learning~\cite{Binder} abilities for desired results.

\textbf{Identifying/Refining insights.} For each round of Q\&A, we instruct the agent to carefully examine and determine whether it contains insights and output an insight list (Figure~\ref{fig:workflow}B1(c)). Meanwhile, we maintain the \textit{previously} extracted insights as the agent's memory (Figure~\ref{fig:workflow}B1(b)), which not only helps it leverage in-context learning to extract and output insights in a consistent manner, but also enables the refinement of previous insights. During conversational data analysis, users may not always pose a new question every time; instead, they often iteratively adjust their prompts for clarification or enhancement~\cite{chopra2023conversational}. For example, a user may request an alternative visualization to better illustrate a particular insight. Therefore, by directing the agent to choose between two actions (\ie \verb|`identify new insight'| or \verb|`refine existing insight'|), we ensure a comprehensive analysis of each round of Q\&A without missing any follow-up information.
Moreover, rather than replicating LLMs' verbose responses, the extracted insights are always \textit{summarized} into concise sentences for intuitive understanding. This eliminates users' burden of extra prompt engineering to retrieve quick answers for simple data queries, while systematically extracting all insights for high-level questions.

\textbf{Associating insight evidence.} To automatically bind all relevant insight evidence with each insight (Figure~\ref{fig:workflow}B1(d)), the agent is required to scrutinize the code, code outputs, visualizations, and NL explanations in LLM responses, focusing on their data and semantic implications.
This allows the agent to locate the \textit{minimum} but \textit{critical} parts that directly support each insight, which mitigates users' manual and cognitive load in understanding and recording insights without having to examine the entire contexts.
We provide in-context examples for each type of insight evidence to improve the agent's awareness and performance of the task. Meanwhile, we instruct the agent to also take previous insights into consideration for potential modifications, in case that new evidence may emerge due to users' iterative prompting.

\textbf{Evaluating insight interestingness.} Inspired by QuickInsights~\cite{10.1145/3299869.3314037}, we judge insight interestingness (\textbf{R3}) by two factors: its \textit{semantic significance} (\ie the subject of it should be important, such as a best-selling product) and \textit{statistical significance} (\ie the relevant statistical metrics of it should be notable, such as a high standard deviation).
\begin{enumerate}[leftmargin=*]
    \vspace{-1mm}
    \item The agent evaluates each insight's semantic meaning and assigns a \textit{semantic score} $S_{sem}$ of 1 to 5 based on its overall understanding of the insight under the analytic context. For instance, if the user focuses on product profit, the 1\textsuperscript{st} most profitable product is more significant than the 3\textsuperscript{rd} one. We instruct the agent to consider multiple aspects (\ie significance, impact, relevance)~\cite{10458347} for a comprehensive assessment, and provide in-context examples (\ie insight-score pairs) and previous scores to enhance scoring performance and consistency.
    \vspace{-1mm}
    \item The agent categorizes the insights and uses function calls to calculate their corresponding statistical metrics (Figure~\ref{fig:workflow}B1(e)). We follow prior works for categorizing insights~\cite{8805442} and mapping insight categories to suitable statistical metrics~\cite{10.1145/3472749.3474792}. As insights may belong to multiple categories, we employ a \textit{majority-vote} strategy~\cite{sun-etal-2023-text} to determine the most prominent one.
    Then, the agent assigns a \textit{statistical score $S_{stat}$} of 1 to 5 based on the calculated metrics and heuristics adopted from~\cite{8440860, 8805442}. For example, a high Pearson correlation coefficient results in a high $S_{stat}$ for correlation insights.
\end{enumerate}
We combine two scores using a weighted average: $S_{final} = S_{sem} \cdot \omega + S_{stat} \cdot (1 - \omega)$, with the weight $\omega$ empirically set to 0.6. Finally, $S_{final}$ is rounded to a scale of 1 to 5 (Figure~\ref{fig:workflow}B1(f)), with a rationale generated by the agent. Higher scores indicate higher insight interestingness.

\subsection{Insight Organization}
\label{sec:im_agent}

To organize insights from multiple aspects on the fly (\textbf{R2, R3}), the \textit{IO Agent} receives the extracted insights (with relevant evidence) and examines their data and semantic characteristics (Figure~\ref{fig:workflow}B2). It is responsible for determining the corresponding data context and analytic topics/subtopics of each insight. Based on the \textit{IO Agent}'s outputs, we sequentially categorize the insights into different subgroups. We introduce our prompt techniques and topic classification method below.

\textbf{Providing overall analysis domain.} To ensure the identification of valid data attributes and relevant analytic topics, we provide an automatically generated NL description of the dataset, its first five rows, and a list of its attributes beforehand (Figure~\ref{fig:workflow}B2(a)). This enables the agent to gain an overall understanding of the current analysis domain.

\textbf{Determining data context.} The agent is tasked with identifying the corresponding data attributes associated with each insight. To mitigate the risk of fabricating non-existent attributes, we explicitly instruct the agent to restrict its selection to the given attribute list. Meanwhile, it is required to identify the analytical actions (\eg \textit{filtering} and \textit{aggregation}, if any) applied to the data subset pertinent to each insight, based on the insight evidence provided. Consequently, we can obtain each insight's data context (Figure~\ref{fig:workflow}B2(b)) to support users' detailed inspection needs.

\textbf{Classifying into topics/subtopics.}
Traditional topic modeling methods (\eg LDA) are limited when handling short and sparse texts~\cite{10386113} like insights. Also, they generate latent topics (\ie collections of words) that lack clear semantic meanings. Inspired by a recent work~\cite{liang2023c5}, we adopt LLMs to sequentially assign \textit{human-readable} topics (\eg a topic named \textit{Climate Analysis}) for each insight (Figure~\ref{fig:workflow}B2(c)).
\begin{enumerate}[leftmargin=*]
    \item First, we maintain a list of current topics that are generated based on previous insights as the agent's memory.
    \vspace{-1mm}
    \item For each new insight, the agent is instructed to \verb|select| a suitable topic from the list that best describes its semantic meaning. To combine LLMs' NL understanding abilities with a best practice from prior literature~\cite{reimers-gurevych-2019-sentence}, we provide cosine similarities between the embeddings of the insight and each existing topic for reference, enabling the agent to make more informed decisions.
    \vspace{-1mm}
    \item In cases where no existing topic semantically describes the insight, or when the topic list is initially empty, the agent must \verb|generate| an appropriate analytic topic by abstracting the insight into a concise and high-level title. We prompt the agent to ensure that the new topic falls within the provided analysis domain and is broad enough to encompass similar subsequent insights. To avoid generating identical or overlapping topics, the agent must utilize function calls to calculate the cosine similarities between the candidate new topic and each existing topic. We empirically set the similarity threshold at 0.55. If any similarity score exceeds this threshold, the agent must generate another new candidate topic. Once the new topic is determined, it is added to the topic list for future selection.
    \vspace{-1mm}
    \item Finally, the selected or generated analytic topic for the newly extracted insight is determined. We then recursively execute the above steps to classify subtopics within the assigned main topic.
\end{enumerate}
Notably, we use the \texttt{all-MiniLM-L6-v1} model from Sentence Transformers~\cite{reimers-gurevych-2019-sentence} for embedding calculation.

\textbf{Identifying related insights.} After obtaining the data context and analytic topics of the extracted insights, we categorize them into subgroups to enable user navigation from different aspects. We also determine related insights across two dimensions (Figure~\ref{fig:workflow}B2(d)). First, we identify \textit{data-related} insights by comparing the intersections between their associated data attributes. For example, an insight associated with \verb|[MPG, Year, Origin]| is closely related to another one associated with \verb|[MPG, Year]|. Second, we identify \textit{semantic-related} insights by comparing the cosine similarities between their embeddings. Consequently, two lists of related insights are derived for each insight (Figure~\ref{fig:workflow}B2(e)). By linking them together, we address the common user need for easier reference or comparison of similar data findings.

\section{InsightLens: User Interface}

Built upon the LLM-agent-based framework, \name \space features a user interface (Figure~\ref{fig:workflow}A) to facilitate insight management and navigation during LLM-powered data analysis.
In this section, we first present an overview of the user interface, and then describe its core features, visual designs, and interactions, including \textit{User Input} (Figure~\ref{fig:workflow}A1), \textit{Insight Inspection} (Figure~\ref{fig:workflow}A2), and \textit{Topic Exploration} (Figure~\ref{fig:workflow}A3).

\subsection{User Interface Overview}
The user interface of \name \space features five coordinated views (Figure~\ref{fig:system}). It is designed to augment existing interfaces while maintaining users' original conversational workflow (\textbf{R4}). Given the unique nature of conversations which display the most information at first glance, we sought advice from data analysts in our formative study and iteratively refined our visual designs. Consequently, we choose to adopt a \textit{`details first, overview last'} strategy~\cite{8440850} from left to right to make the user interface more applicable to the conversational workflow, while facilitating easy inspection and navigation of insights during analysis.

To achieve this, we keep the \textit{Chat Window} (Figure~\ref{fig:system}A) similar to ChatGPT on the left, where users can input their analytic intents and view LLMs' responses. Beside it, the \textit{Insight Details} (Figure~\ref{fig:system}B) shows an individual insight with its relevant data context and supporting evidence for thorough inspection, while the \textit{Insight Gallery} (Figure~\ref{fig:system}C) displays its data- and semantic-related insights for convenient comparison. Additionally, we employ a matrix-based design in the \textit{Insight Minimap} (Figure~\ref{fig:system}D) to chronologically visualize the analysis process. Each row represents a unique insight, showcasing its data and semantic characteristics. Finally, the \textit{Topic Canvas} (Figure~\ref{fig:system}E) on the right adopts a tree-based design to visualize the hierarchical topic structure, enabling users to explore their findings across different analytic topics.

\subsection{User Input \& Insight Inspection}

As the entry point of the user interface, users upload their datasets and interact with the analytical chatbot in the \textit{Chat Window}.
Right beside it lays the \textit{Insight Details} and \textit{Insight Gallery} arranged vertically to enable detailed inspection for each insight.
Along with the conversation flow, we provide an overview of all extracted insights in the \textit{Insight Minimap}, which is constructed by \textit{insight rows} vertically stacked in temporal order. These four views are coordinated to \textit{scroll together} seamlessly. By clicking on each insight row, users can conveniently examine its details and navigate between different parts of the conversation. Collectively, these progressively-evolving visualizations support the following tasks to facilitate multi-level and multi-faceted insight navigation without disrupting the conversational workflow (\textbf{R3, R4}).

\textbf{Inspecting insight details.} As the conversation progresses, the \textit{Insight Details} updates with the latest extracted insight. It consists of five sections  (\ie \textit{Data}, \textit{Code}, \textit{Code Output}, \textit{Vis}, and \textit{Insight}) to display the insight's summary along with its associated data context and evidence. These sections are \textit{collapsible} to enable details on demand and to satisfy different user background and preferences (\eg some analysts might be unfamiliar with coding and prefer to view only the data context or visualizations). By default, the \textit{Code} and \textit{Code Output} sections are collapsed to benefit non-technical users. Meanwhile, the relevant NL explanations are \textit{highlighted} in LLMs' original responses in the \textit{Chat Window}. All these content are the \textit{minimum} but \textit{critical} parts of the intermediate outputs to reduce users' cognitive load. To navigate among different insights, users can either 1) scroll in the \textit{Chat Window} or \textit{Insight Minimap} or 2) click on the dots (\icon{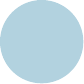}) below each response. Pinning (\icon{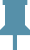}) is also supported to temporarily disable scrolling coordination to focus on a specific insight.

\textbf{Comparing related insights.} In accordance with the currently focused insight of the \textit{Insight Details}, we present its related insights in the \textit{Insight Gallery}, ranked by similarity (or by temporal order for ties). For simplicity, only the associated visualization and the insight's summary are displayed in each \textit{insight card}. To enable a clear understanding of the rationales behind each recommendation, we show the relevant data attributes for data-related insights and similarity scores for semantic-related insights. Users can click on each insight card in the gallery to view its details for comparison or reference.

\textbf{Revealing data coverage.} On top of the minimap, we provide a histogram (Figure~\ref{fig:system}D1) to visualize the distribution of the associated insight counts across each data attribute. By observing the histogram, users can intuitively understand which attributes have already been extensively analyzed and which ones remain underexplored. Hovering and sorting are also supported to view detailed information and quickly locate the uncovered attributes.
Therefore, users' awareness of their data coverage during analysis can significantly be improved.

\textbf{Understanding context transitions.} In each insight row of the minimap (Figure~\ref{fig:system}D2), we represent its associated data attributes with a set of connected points (corresponding to the above histogram). The horizontal connecting lines can visually indicate the holistic consideration of the involved attributes in each round of Q\&A. We also provide vertical reference lines activated by hovering over any point to maintain alignment with the histogram. These insight rows not only enable a quick review of each insight's data context, but also showcase context transitions throughout the analysis process, which reveal the change of users' focused attributes. For example, certain visual patterns can represent different types of transitions like \textit{continue} (\icon{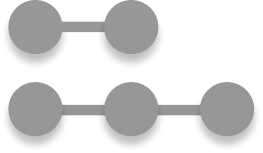}), \textit{retain} (\icon{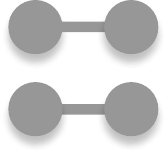}), and \textit{shift} (\icon{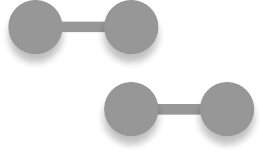})~\cite{10.1145/3472749.3474792}. Grasping these transitions helps users track the progress of analytic conversations~\cite{8019833}, thereby mitigating the risk of analytical biases, such as focusing excessively on specific data subsets.
In case that users expect to prioritize some attributes of interest, \eg always monitoring `\verb|Worwide Gross|' for financial analysis, they can \textit{drag} the bars in the above histogram to adjust column order. Additionally, we colorize each insight row to denote its analytic topic and reveal the topic changes. Overall, this simple and intuitive design can be seamlessly integrated into the conversational workflow and helps users better review their analyses across both data and semantic dimensions.

\textbf{Highlighting insight interestingness.} To empower users to easily identify and revisit high-quality or interesting insights, we visualize the interestingness scores of each insight as horizontal bars (Figure~\ref{fig:system}D3), as well as adding a category tag in each insight row for reference. As the `interestingness' of an insight can be subjective and varies among users~\cite{8440860}, the scores automatically assigned by LLMs may not accurately reflect user preferences (\ie whether they would find the insight significant). To balance this, we provide LLMs' explanations for the rationales behind each interestingness score on hovering, and also allow users to dynamically \textit{adjust} the score by resizing the corresponding bar. Therefore, this feature offers an alternative way for users to manage and navigate previous insights, either based on automated evaluations or their own judgment, similar to a `bookmark' for insight significance.

\subsection{Topic Exploration}

As the highest-level overview, the \textit{Topic Canvas} visualizes the hierarchical topic structure of all extracted insights. We choose the tree-based design due to its simplicity and intuitiveness for topic organization and exploration (\textbf{R3, R4}). The tree (without a root node) is structured into two levels, representing main topics and their subtopics, respectively. Each node indicates a topic/subtopic, differentiated by color and labeled with its title and associated insight count. These nodes are visually linked to their corresponding insight rows in the \textit{Insight Minimap}. Additionally, hovering over any node will highlight its included insights (and subtopics, if any) and display a brief description for quick inspection of each topic's essence. Overall, the \textit{Topic Canvas} is automatically updated along with the analysis process and coordinated with other views to facilitate insight navigation across analytic topics.
\section{Technical Evaluation}

The effectiveness of \name \space depends on whether our framework can successfully record and organize the LLM-generated insights. Therefore, we conducted a technical evaluation focusing on (1) the \textit{coverage} of insight extraction, (2) the \textit{accuracy} of insight evidence association, and (3) the \textit{quality} and \textit{accuracy} of insight organization.

\subsection{Experiment Settings}

\textbf{Dataset.}
We collected 10 datasets from reputable sources (6 from Kaggle and 4 from Vega) with diverse analysis domains (\eg education, economics) and number of rows (\(\mu=1058, \sigma=777\)) and columns (\(\mu=14, \sigma=5\)). We manually crafted 10 analytic queries for each dataset, totaling to 100 samples. These queries, together with their corresponding datasets, were input into our system, resulting in 104 extracted insights and 50 generated analytic topics (with 70 subtopics).

\textbf{Methodology.}
To evaluate insight extraction, we first manually identified and labeled the key insights in the original responses generated by the analytical chatbot, providing a ground truth for the insights extracted by the \textit{IE Agent}.
Then, we measured the ratio of covered labeled insights to their total number (\ie coverage). As the automatically extracted insights were summarized by the \textit{IE Agent} for easier understanding, we considered a labeled insight as covered if its semantic meaning was \textit{contained} in the corresponding extracted insight.

To evaluate evidence association, we measured the ratio of insights with correctly associated evidence to the total number of extracted insights (\ie accuracy). If any part of the evidence (\ie code, code outputs, visualizations, and NL explanations) was incorrect or irrelevant to its corresponding insight, we considered it as a negative sample.
To ensure robustness, we adopted a four-step verification process: (1) confirming the exact match between the associated evidence and the original response; (2) evaluating the correctness of data processing through manual code review and execution results; (3) validating the appropriateness of visualizations based on the ground truth processed data; and (4) examining the relevance of the associated evidence to the corresponding insight through manual assessment.

To evaluate insight organization, we focused on two aspects: data and semantic characteristics (see Section~\ref{sec:im_agent}). For data context, we measured the ratio of insights with correctly identified data attributes (and analytical actions, if any) to the total number of extracted insights (\ie accuracy). For analytic topics/subtopics, we utilized GPT-4 to rate their quality, a widely adopted method in the NLP community for assessing machine-generated texts that has proven effective in various scenarios~\cite{chiang-lee-2023-large, liu-etal-2023-g, gao2024llmbasednlgevaluationcurrent}. Specifically, we instructed GPT-4 to consider multiple aspects of the topics (\eg relevance, clarity, adaptability) for a thorough evaluation. The detailed prompts can be found in the supplemental material.
As the assignment of analytic topics is subjective and lacks a definitive ground truth, we compared the rating scores of our dynamically generated topics with a static baseline~\cite{liang2023c5} (\ie feeding all insights to GPT-4 for topic generation). We then manually labeled each insight with the topic list generated by our system as a ground truth for evaluating topic classification accuracy.

\subsection{Results}

\textbf{Metrics.}
For insight extraction, the coverage of the extracted insights was \textbf{91.2\%} (\ie covered 176 out of 193 labeled insights). For {evidence} association, the accuracy of the associated insight evidence was \textbf{88.5\%} (\ie 92 corrects and 12 errors). For insight organization, the accuracy of the identified data context was \textbf{88.5\%} (\ie 92 corrects and 12 errors). Additionally, analytic topics generated by our system received an average quality rating of \textbf{7.6} on a 10-point scale, surpassing the static baseline (5.9). The accuracy of topic classification was \textbf{91.3\%} (\ie 95 corrects and 9 errors). Overall, these statistical metrics demonstrated the effectiveness and robustness of our LLM-agent-based framework.

\textbf{Failure Cases Analysis.}
For insight extraction, we categorized the 17 failure cases into two types: (1) \textit{Missing Insights} (8/17) and (2) \textit{Missing Details} (9/17). The \textit{IE Agent} sometimes failed to extract all the key insights; instead, it tended to only focus on the most significant ones. For instance, with the query \textit{`compute the average discount percentage offered by each smartphone brand'}, only the brands with the highest and lowest discounts were highlighted, while the {analytical chatbot} actually mentioned numerous intermediate brands in its response. In other cases, the agent over-summarized the information, omitting critical details. An example of this is an extracted insight that merely acknowledged the \textit{`top 10 most profitable movies'} without specifying their titles.

For {evidence} association, we observed two failure modes: (1) \textit{No Code/Code Output} (5/12) and (2) \textit{Incorrect NL Explanations} (7/12). In the former, the \textit{IE Agent} did not include any associated code or code output in its responses. In the latter, it provided incorrect NL explanations that did not align with the insights, arising from either fabricated sentences or an oversimplification of the original output.

For insight organization, we evaluated failures in terms of data context accuracy and topic classification accuracy. Data context errors primarily stemmed from \textit{Fabricating Attributes} (9/12), with the remainder due to \textit{Missing Attributes} (3/12). The former occurred when the {analytical chatbot} created new attributes for specific queries (\eg defining a \verb|Decade| attribute from \verb|Year|), leading to the \textit{IO Agent}'s inability to correctly identify the original dataset attributes. In contrast, the latter was due to the agent's occasional failure to fully deduce the associated attributes. Regarding topic classification, the predominant issue was \textit{Topic Disagreement} (9/9), where humans and {GPT-4} focused on different aspects. Since insights could span multiple topics, such cases were technically not `errors' but rather outcomes of varying labeling criteria.

Overall, most failure cases discussed above can be ascribed to LLMs' hallucinations. Such issues are particularly evident given the intricate nature of our targeted tasks and the complex prompting techniques we employ for our framework, which often lead to LLMs' generation of unexpected outputs. To mitigate this, we can incorporate more effective instructions to make LLMs' behavior more reliable and robust~\cite{zhang2023sirens}.

\textbf{Summary.}
Despite the few failure cases, the results demonstrated our framework's high coverage, accuracy, and quality in automated insight recording and organization. This can significantly reduce users' manual and cognitive effort during conversational data analysis, establishing a solid foundation for the interactive features of \name.

\section{User Study}

To evaluate the effectiveness of \name \space in facilitating insight {management and navigation} during LLM-powered data analysis, we conducted a within-subjects user study. Specifically, we aimed to collect users' feedback on the effectiveness and usability of \name's features, as well as its impact on the overall data analysis process.

\subsection{Experiment Design}

\textbf{Participants and Setup.}
We recruited 12 data analysts (P1-12, 4 females and 8 males, age from 24 to 29) from the business intelligence department of a local technology company. {Their expertise levels in data analysis ranged from junior/medium (8/12, \verb|<| 5-year experience) to senior (4/12, \verb|>| 5-year experience).}
Their daily tasks included analyzing datasets and reporting data findings, with proficiency in various tools like Excel (12/12), Python (10/12), and Microsoft Power BI (8/12). All of them had experience using LLMs (\eg ChatGPT, Claude, Qwen) for their work with varying frequencies (6 often, 4 sometimes, 2 rarely). Each participant received \$25 as compensation upon completion.

\name's visual support for insight management and navigation primarily relies on the four coordinated views (\ie \textit{Insight Details}, \textit{Insight Gallery}, \textit{Insight Minimap}, and \textit{Topic Canvas}) to function as a whole. Therefore, we set the comparative \textit{Baseline} as the \textit{Chat Window} of \name \space excluding all interactive features to evaluate their effects, similar to prior studies on LLM data analysis interfaces~\cite{waitgpt}. This ChatGPT-like \textit{Baseline} mirrored the systems familiar to participants for LLM-powered data analysis and maintained the same appearance and chat functionality as \name \space for a fair comparison. We also provided a document editor for participants to record their findings.

\textbf{Tasks and Datasets.} Participants were asked to use both \name \space and \textit{Baseline} to analyze two datasets: (1) a housing dataset (15 columns, 1460 rows) and (2) a colleges dataset (14 columns, 1214 rows). They were instructed to perform open-ended data exploration on each dataset to provide insights into (1) the housing market dynamics for real estate agents, and (2) the various factors of US colleges for student applicants, as if they were to provide a comprehensive data report within a week. To mitigate learning effects while ensuring comparability of collected data across different experiment sessions, we split each dataset into two parts~\cite{DBLP:conf/iui/LiLTWPMC21}, each of which was allocated to one of the systems.

\textbf{Procedure.} Initially, participants were asked to sign a consent form and fill out a pre-study questionnaire to collect their demographic information. After that, we conducted a tutorial using an example dataset to introduce the features of both systems. Participants were then given adequate time to familiarize themselves with each system, during which they were encouraged to raise any questions or concerns.

Then, participants were requested to use both systems across two datasets (and tasks). We counterbalanced the order of the systems and datasets (4=2x2 sessions in total) to mitigate learning effects. Each session lasted 15 minutes and was screen- and audio-recorded as system logs. Participants were also encouraged to think aloud about their thoughts and findings during the analysis process.

Finally, participants were required to complete a post-study questionnaire using a 5-point Likert scale, followed by a semi-structured interview to comprehend their ratings and collect qualitative feedback on the effectiveness, usability, and potential impact of the system on their daily workflow. The entire study lasted about 120 minutes.

\textbf{Measures.} We collected 48(=12x4) recordings and system logs. To complement participants' qualitative feedback, we employed the following measures: (1) \textit{number of {recorded} insights}, (2) \textit{number of unique data attributes explored}, and (3) \textit{number of unique analytic topics explored}. These measures were informed by previous literature~\cite{10301695, 9552196} and offered quantitative evidence for our analysis. To ensure methodological consistency, we utilized the same prompting techniques of \name \space on \textit{Baseline} for data processing.

\subsection{Results}

\begin{figure}[t]
  \centering
  \includegraphics[width=\linewidth]{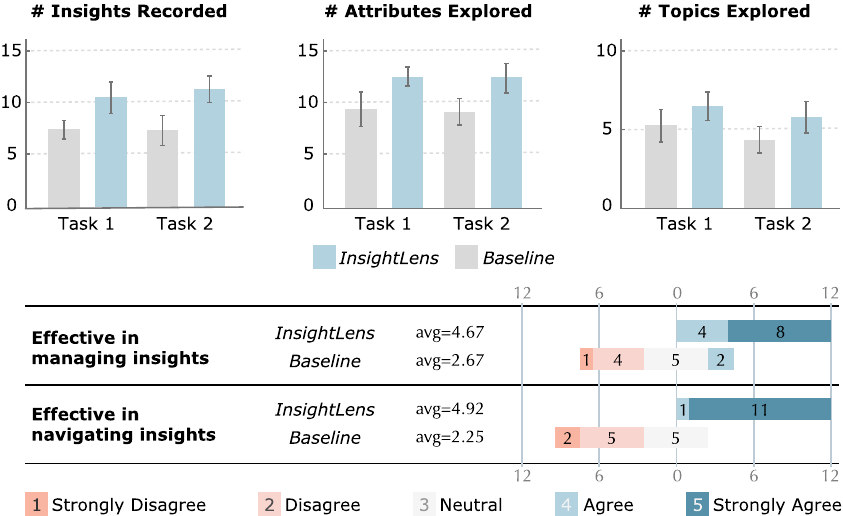}
  \caption{The results of the measures and qualitative ratings regarding \name's support for insight {management and navigation}.}
  \label{fig:ratings1}
  \vspace{-3mm}
\end{figure}

\begin{figure}[t]
  \centering
  \includegraphics[width=\linewidth]{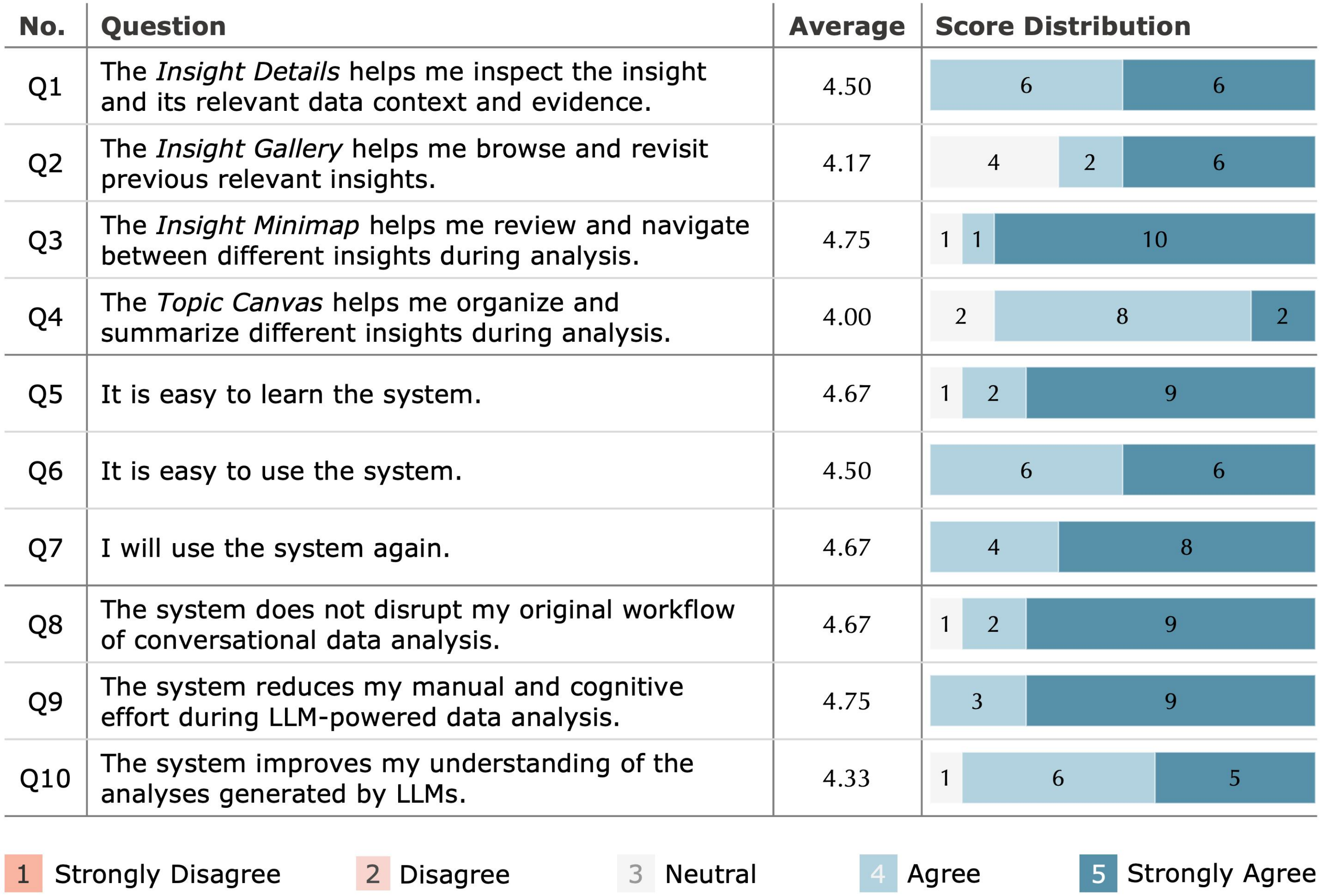}
  \caption{The results of the questionnaire regarding \name's effectiveness, usability, and impact on data analysis.}
  \label{fig:ratings2}
  \vspace{-6.5mm}
\end{figure}

All participants completed four experiment sessions successfully. Based on their qualitative feedback and the collected quantitative measures, we discuss the effectiveness of \name \space in facilitating insight {management and navigation} (Figure~\ref{fig:ratings1}). We then report \name's feature effectiveness, system usability, and impact on data analysis (Figure~\ref{fig:ratings2}).

\textbf{Support for Insight {Management}.}
The effectiveness of \name \space in facilitating insight {management} was appreciated by all participants (\(\mu=4.67>2.67, p=.002\)).
{
Recording insights was much easier in \name, whereas \textit{Baseline} forced participants to manually scrutinize and summarize LLMs' lengthy responses. P3 expressed his favor for \textit{`the dots below each message'} that \textit{`reminded him of missed out insights'}. We also observed that participants constantly referred to the \textit{Insight Details} to review and record the relevant insight evidence, which allowed them to \textit{`easily see the involved attributes and charts without scrolling up and down'} (P10). For organizing insights, the progressively updating \textit{Topic Canvas} and \textit{Insight Minimap} significantly eased participants' burden, mitigating the need for \textit{`resorting to tools like Word'} (P5) and \textit{`summarizing an insight outline'} (P6).
}

Additionally, one of our measures reinforced \name's support for insight {recording}. Specifically, participants {recorded} more insights using \name \space compared to \textit{Baseline} (Task 1: \(\mu=10.4>7.4, p=.002\); Task 2: \(\mu=11.1>7.3, p=.005\)). We ascribed the observed significant difference to \name's support for reducing the time needed for locating insights and their relevant evidence, thereby leading to more insights recorded within a limited time frame.

\textbf{Support for Insight Navigation.}
\name \space was rated as highly effective in reviewing and navigating previous insights (\(\mu=4.92>2.25, p=.002\)). Participants highly valued \name's features for {navigating} insights from different aspects. For example, P4 appreciated \textit{`tracking her findings by time order in the minimap'}, while \textit{`using the baseline required her to {scroll} back and forth to grasp what she explored before'}. During open-ended data exploration, participants recognized the importance of maintaining awareness of the overall analysis flow, which avoided \textit{`repetitive analyses on previously explored topics'} (P8).

Interestingly, the quantitative measures revealed the potential expansion on participants' data and analytic coverage due to their improved awareness of the analyses. When using \name, they explored more data attributes (Task 1: \(\mu=12.4>9.3, p=.006\); Task 2: \(\mu=12.3>9.1, p=.012\)) and analytic topics (Task 1: \(\mu=6.5>5.3, p=.03\); Task 2: \(\mu=5.8>4.3, p=.035\)) than \textit{Baseline}. During the experiments, we constantly noticed that participants checked and navigated in the \textit{Insight Minimap} or \textit{Topic Canvas} before posing their next query. Consequently, these observed significant differences implied participants' tendency to analyze more comprehensively when {provided with easier navigation of the recorded} insights organized across data and semantic dimensions.

\textbf{Feature Effectiveness.} Overall, the features of \name \space were well-received by most participants.
Firstly, the \textit{Insight Details} (Q1) was appreciated by participants for allowing them to \textit{`quickly obtain an insight summary without manually reading every piece of messages'} (P5, P7). Also, the associated insight evidence such as code snippets eliminated their need to \textit{`scroll back to find that specific line of code for data transformation'} (P6) to {comprehensively record the insight.}
Secondly, the \textit{Insight Gallery} (Q2) helped participants review related insights conveniently. P8 found it particularly useful for \textit{`understanding attribute relationships when dealing with multiple similar insights'}, while P3 likened it to \textit{`a menu tool'} that enabled him to review different visualization types for similar insights. However, some participants found it less beneficial (P2, P4) due to the rather short analysis time allotted for the experiments.
Thirdly, the \textit{Insight Minimap} (Q3) was constantly praised by most participants (8/12) as \textit{`the most useful feature'} (P1).
P9 described it as \textit{`being very innovative and reminded him of the minimap in VS Code'}, while others favored its \textit{`clear presentation of covered data attributes'} (P2, P4, P7, P12) and \textit{`color encodings to reveal topic changes'} (P5). This made the analysis process \textit{`more structured and thorough'} (P11). Additionally, the interestingness bars enabled participants to discard trivial insights. For example, P4 identified an insight with an extremely low interestingness score about a negligible attribute relationship \textit{`caused by an accidental query'}.
Finally, the \textit{Topic Canvas} (Q4) reduced participants' manual and cognitive effort to organize insights. The generated topics were reported as \textit{`being reasonable and intuitive'} that \textit{`decreased the chaos of the overwhelming conversation'} (P10).
Moreover, viewing the tree-based topic structure gave P3 a sense of \textit{`solving the open-ended task from various angles'} - aiding comprehensive thinking - though some preferred relying on personal judgment rather than \textit{`being disturbed by the organized topics'} (P5).

\textbf{System Usability.} All participants found \name \space easy to learn (Q5) and use (Q6), and were willing to integrate it into their daily workflow (Q7). The visual designs were praised as \textit{`very intuitive and user-friendly'} (P3, P7) without \textit{`causing steep learning curves'} (P1). P9 noted that the views looked so natural that \textit{`any professionals could understand its main features at first glance'}. Meanwhile, participants also suggested improvements for \name. For example, P4 {wished to intervene the organization process by \textit{`proposing her own topics'}}, and P11 expected to \textit{`combine certain insights for more in-depth analysis'}.

\textbf{Impact on Data Analysis.} We examined \name's impact on LLM-powered data analysis workflows for fluidity, workload, and understanding. Firstly, participants agreed that \name \space was unobtrusive and did not disrupt their conversational interactions (Q8). P9 noted, \textit{`he just chatted with LLMs as usual, and the views updated automatically without interference'}, while P7 described it as \textit{`essentially a chat interface augmented with useful plugins'}. Secondly, \name \space reduced manual and cognitive load (Q9), alleviating issues like \textit{`excessive scrolling'} (P2) and \textit{`memorizing insights in mind'} (P12). Recording and organizing insights on the fly helped participants \textit{`focus more on the analysis itself rather than constant context switching'} (P10). Finally, \name \space improved participants' understanding of LLM-generated analyses (Q10). P6 remarked, \textit{`it felt like she was more involved in the analysis process by inspecting progressive changes in views, instead of merely inputting queries and waiting for LLMs to handle everything'}. Thus, \name \space helped strike a balance between automation and human agency, thereby increasing users' understanding and engagement.

\subsection{Observed Behaviors}

We observed two prominent workflow patterns adopted by different participants when using \name \space for data analysis.

\textbf{User-Initiated Workflow.} Participants with a clear analysis goal often posed sequential queries based on their own judgment and preferences with minimal system intervention. For example, P5 explored the colleges dataset focusing on how college ownership influenced factors like student quality and financial condition. Here, the \textit{Insight Minimap} and \textit{Topic Canvas} primarily served as structured and organized ways for reviewing previous insights rather than inspiring new discoveries. The construction of the topic tree mainly progressed from \textit{bottom} (insights) to \textit{top} (topics) with more subtopics than main topics, revealing a depth-oriented exploration pattern.

\textbf{System-Initiated Workflow.} Participants without a specific aim, often due to unfamiliarity with the analysis domain, initially posed multiple random queries to \textit{`make a draft'} (P1). They then inspected the \textit{Insight Minimap} and \textit{Topic Canvas} to gain an overview of their analyses and observe potential biases (\eg certain attributes/topics may have been thoroughly explored while others remain overlooked) to plan future explorations. Therefore, the construction of the topic tree was now from \textit{top} to \textit{bottom} with many topics scattered around and few subtopics, reflecting a breadth-oriented exploration pattern.
\section{Discussion}


\subsection{Design Implications}

\textbf{Integrate data and semantic context for enhanced understanding.}
Given the limitations of linear chat-based interfaces, managing LLMs' contexts for complex tasks has gained popularity in VIS and HCI~\cite{10.1145/3586183.3606737, 10.1145/3586183.3606756, liang2023c5}. Unlike existing works that primarily extract semantic structures, \name \space further integrates data context - crucial for data analysis - including data attributes and analytical actions. We visualize dynamic data and semantic context simultaneously in a minimap, allowing users to quickly grasp the analysis process. Our user study shows that this integration not only aids in reviewing and navigating insights but also potentially expands data analysts' data and analytic coverage, leading to more comprehensive results in exploratory data analysis.

\textbf{Provide follow-up analytic guidance for data exploration.}
In our user study, many participants (6/12) suggested incorporating query recommendations, particularly for unfamiliar datasets (\ie the \textit{`cold start'} issue).
Prior research has extensively explored analytic guidance~\cite{wang2022interactive, 10.1145/3472749.3474792}, which can be improved with LLMs' capabilities~\cite{gu2023data}. \name's support for organizing insights on the fly can establish a robust foundation for context-aware assistance. For example, integrating another agent into our framework can generate tailored suggestions by considering analysts' background, goals, and current focused topics and attributes, thereby deepening or broadening their analyses.

\textbf{Balance between flexibility and complexity of interaction paradigms.}
Our design principle maintains a conversational workflow primarily through natural language. However, we recognize the potential of other modalities for NLI-based data analysis (\eg direct manipulations~\cite{8019860} and sticky cells~\cite{wangStickyLandBreaking2022}).
For example, some participants in our user study expected to modify the \textit{Topic Canvas} by adding or editing nodes, akin to mind maps. While such features could enhance LLM interaction flexibility~\cite{10.1145/3586183.3606756}, they may also introduce complexities~\cite{10.1145/2678025.2701376}.
Therefore, we aim to achieve a trade-off between NLIs' intuitiveness and visualizations' expressiveness. Future research could further explore how to balance these aspects in designing LLM interaction paradigms for data analysis.

\subsection{More Application Scenarios}
Incorporating LLMs into data analysis is an emerging but promising paradigm. With LLMs' growing reasoning capabilities and extended context windows~\cite{chen2023extending}, data analysts can potentially conduct longer and more in-depth analyses on intricate datasets. Such envisions necessitate smart strategies to manage complex analytic contexts.
While \name \space focuses on augmenting conversational interfaces, its design rationales can be adapted to traditional data analysis workspaces, such as BI platforms~\cite{powerbi} and computational notebooks~\cite{10296056}. For example, in BI platforms like Tableau~\cite{tableau}, users conduct visual analysis through a dashboard and a sidebar-based analytical assistant. An \textit{Insight Minimap} can serve as an analytic timeline, allowing users to revisit previous visualization states and maintain awareness of the entire analysis process. Additionally, within Jupyter Notebooks, users typically interact with LLMs via code comments or magic commands~\cite{10.1145/3544548.3580940} to perform exploratory data analysis. Constructing a \textit{Topic Canvas} that reflects the semantics of code and markdown text can enable a non-linear, tree-based navigation of notebook cells, offering an innovative way to organize messy notebooks.
Therefore, we believe that our work could inspire future research in making LLM-powered data analysis more streamlined, accessible, and productive through visualizations.


\subsection{Limitations and Future Work}
\label{sec:limit}

\textbf{Hallucination.}
LLMs can generate incorrect or misleading insights~\cite{gu2023analysts}. While our main focus is on insight management and navigation, \name \space can inherently support some degree of verification by displaying relevant evidence like code snippets or outputs, allowing users to identify potential errors.
For example, during our user study, participants frequently reviewed the \textit{Data} section within the \textit{Insight Details} to verify data attributes and analytical actions. This helped them quickly determine if LLMs had correctly utilized and transformed the data, without tediously sifting through lengthy responses.
To further improve \name's reliability, we can fine-tune LLMs for data analysis tasks ~\cite{zhang2023sirens} and explore advanced agent designs~\cite{hong2024datainterpreterllmagent}.
Moreover, incorporating methods like code verification~\cite{waitgpt} and task decomposition~\cite{10.1145/3654777.3676345} can help users proactively diagnose issues in LLM responses.
We can also examine data transformation errors and factual contradictions via self-reflection~\cite{ji-etal-2023-towards} and external knowledge bases~\cite{10297594}, and highlight them with visual cues like warning icons~\cite{10.1145/3613904.3641904}.

\textbf{Bias.}
\name \space relies on LLMs for fully automated insight extraction and topic generation. The lack of interactive control may foster an excessive reliance on LLM outputs, potentially introducing biases~\cite{10.1145/3460231.3474244}. Users might overlook critical insights requiring human interpretation or feel constrained by LLM-generated topics.
Moreover, biases inherent in LLMs' training corpus can be exacerbated by the patterns of the current dataset, which may further propagate into the extracted insights or generated topics~\cite{10.1162/coli_a_00524}.
To mitigate these issues, we can integrate visual alert mechanisms (\eg highlighting uncertainties in LLM responses~\cite{9973204}) to improve user awareness of potential biases. 
Additionally, incorporating user feedback loops, such as enabling adjustments or merging of topic nodes~\cite{10.1145/3586183.3606756}, can allow users to provide background knowledge or specify personal preferences. This can enhance the customization of insight extraction and topic generation.

\textbf{Scalability.}
To handle complex queries and large numbers of insights/topics, \name \space employs a streaming strategy~\cite{xie2023openagents} to enable real-time parsing and rendering of LLM responses, thereby maintaining system responsiveness. This allows for immediate presentation of processed insights/topics and dynamic updating of visualizations to significantly reduce user wait times. Nevertheless, we recognize that it requires additional engineering efforts to support vast numbers of attributes or topics in large-scale data analysis scenarios.
Currently, the system's response time is primarily affected by the LLM inference latency, which can be mitigated through hardware acceleration solutions such as Groq~\cite{10.1145/3470496.3527405}.
To address potential visual clutter in the user interface, 
hierarchical semantic zoom~\cite{suh2023structured} for nodes within the \textit{Topic Canvas} and adaptive visual filtering~\cite{5204083} for columns in the \textit{Insight Minimap} can enhance user interaction and prevent interface overload.

\textbf{Cognitive Impact.}
\name \space utilizes color encoding to differentiate analytic topics, which may introduce cognitive trade-offs like color fatigue~\cite{7192718} or visual clutter, especially with extensive use or among users with color vision deficiencies. Future work should explore alternative salience strategies to complement color encoding and evaluate how different visual design choices might affect \name's effectiveness. For instance, incorporating additional visual encodings (\eg size, shape) or multimodal cues~\cite{10670075} can diversify highlighting methods and improve system accessibility.

\textbf{Design and Evaluation.}
The participant groups in our formative and user studies were rather small and lacked diversity in age, gender, and domain, potentially introducing biases and limiting representativeness. A larger, more diverse participant pool would enhance the robustness of our design and evaluation. Additionally, conducting a between-subject study and an ablation study on different user interface components would mitigate individual effects and provide a thorough assessment.

\section{Conclusion}

This work presents \name, an interactive system that visualizes the complex conversational contexts during LLM-powered data analysis to facilitate insight {management and navigation}. Built on an {LLM-agent-based} framework that automates the  {recording and organization} of insights in analytic conversations, \name \space provides a set of {progressively-evolving} visualizations to enable multi-level and multi-faceted {insight navigation}. A technical evaluation and a user study demonstrate the effectiveness of our framework and system.

\bibliographystyle{abbrv-doi-hyperref}

\newpage
\bibliography{template}

\end{document}